%% file: manuscript.tex
\documentclass
[twocolumn, 
singlespacing, 
times
]{aastex631}

\input{commands}

\usepackage{booktabs}

\newcommand{\rev}[1]{\textcolor{black}{#1}}

\shorttitle{Time-domain asteroseismology}
\shortauthors{Dan$^3$ et. al}

\graphicspath{{./}{figures/}}

\begin{document}

\title{Precise Time-Domain Asteroseismology and a Revised Target List for TESS Solar-Like Oscillators} 
\input{authors}

\begin{abstract}

    The \tess\ mission has provided a wealth of asteroseismic data for solar-like oscillators. However, these data are subject to varying cadences, large gaps, and unequal sampling, which complicates analysis in the frequency domain. One solution is to model the oscillations in the time domain by treating them as stochastically damped simple harmonic oscillators through a linear combination of Gaussian Process kernels. We demonstrate this method on the well-studied subgiant star \nuind\ 
    and a sample of Kepler red giant stars observed by \tess, finding that the time domain model achieves an almost \rev{two-fold increase in accuracy for measuring \numax\ compared to typical frequency domain methods}.
    To apply the method to new detections, we use stellar parameters from Gaia DR3 and the \tess\ input catalog to calculate revised asteroseismic detection probabilities for all \tess\ input catalog targets with $T<12$\,mag and a predicted \numax\ $>240\mu$Hz. We also provide a software tool to calculate detection probabilities for any target of interest. 
    Using the updated detection probabilities we show that time-domain asteroseismology is sensitive enough to recover marginal detections, which may explain the current small number of  frequency-based detections of \tess\ oscillations compared to pre-flight expectations.

\end{abstract}


\section{Introduction}

Asteroseismology, the study of stellar pulsations, is a powerful tool for determining the internal and fundamental properties of stars \citep{Christensen-Dalsgaard1983Radiative}. Over the past two decades, the availability of precise and continuous photometry from space missions (such as CoRoT, Kepler, K2) has enabled the characterization of thousands of red-giant and main-sequence stars throughout our galaxy \citep{Appourchaux2008CoRoT,Chaplin2015Asteroseismology,Zinn2020Local, Zinn2022K2}. These measurements are of fundamental importance for the characterization of exoplanets \citep{Huber2013Fundamental} and the study of stellar populations through galactic archaeology (e.g., \citealp{SilvaAguirre2015Ages, Zinn2020Local, Sharma2016LAR, Sharma2019K2HERMES}).

The Transiting Exoplanet Survey Satellite (\tess; \citealp{Ricker2014Transiting}) has continued the space-based asteroseismology revolution, including the detection of oscillations in nearby, bright Sun-like stars \citep{Chontos2021TESS, Metcalfe2020Evolution, Huber202220a} and tens of thousands of red giant stars \citep{Mackereth2021Prospects, Hon2021Quick}. However, initial yields of detected oscillations for stars observed in 2-minute and 20-second cadence \citep{Hatt2023Catalogue} fell below expectations from pre-flight calculations \citep{Schofield2019Asteroseismic}. Possible explanations for lower yields include systematic errors in assumed stellar parameters (which enter into the prediction of detection probabilities) or physical mechanisms such as the suppression of oscillation amplitudes due to stellar activity \citep{Chaplin2011Ensemble}.

\tess\ has surveyed the majority of the sky in a variety of different cadences, duty cycles, and observing strategies. To maximize sky coverage, observations are typically made in noncontiguous sectors of around 27 days, at one ecliptic hemisphere per year. However, large gaps in the time series lead to a poor window function that complicates frequency domain analyses. Several techniques have been developed to overcome this problem, such as separately evaluating the contiguous regions of the light curve, interpolating missing points in the gap (inpainting; \citealp{Pires2015Gap, Pascual-Granado2015MIARMA}), and in some cases removing the gap entirely \citep{Hekker2010Oscillation, Nielsen2022Probabilistic}. The \TESS\ data poses a unique problem in this regard; the majority of observed variable stars now have observational gaps that exceed a year in length, which makes such methods redundant or unwieldy. \citet{Bedding2022Dealing} argue that applying these methods can negatively impact the data, introducing offsets to the measured frequencies and thus lead to an incorrect estimate of stellar properties. With the \tess\ mission delivering on the order of $10^5$ detections for red-giant branch stars \citep{Hon2021Quick}, a new approach is required to determine the global asteroseismic quantities.

Stars with convective envelopes exhibit pressure (p) mode oscillations due to turbulent convection, where repeated sequences of stochastic excitation and damping by motion in the convective layers lead to a range of resonant modes. 
Since oscillations are stochastic, there is no deterministic, parametric model that can fully describe their time evolution. As a result, the analysis of solar-like oscillations is almost always carried out in the frequency domain, where individual excited modes are modeled as Lorentzian peaks \citep[e.g.][]{Kallinger2014Connection}.
One approach, introduced by \citet{Brewer2009Gaussian}, is that of a time-domain formulation, where the stellar oscillations are modeled as a series of damped simple harmonic oscillators through a Gaussian Process (GP; \citealt{Rasmussen2005Gaussian}). 

Gaussian Processes are non-parametric models capable of describing correlated stochastic signals and noise \citep{Aigrain2023Gaussian}. They describe each point as a correlated random variable with a mean value and a variance, where any finite collection of those variables has a multivariate Gaussian distribution. The measure of similarity, or correlation, between pairs of points in the signal to the distance between them in time is given by the covariance function (kernel) of the GP. With a suitable choice of kernel, GPs can describe a wide variety of astrophysical processes, such as quasi-periodic stellar rotation \citep{Angus2018Inferring}, correcting systematics in photometry \citep{Gibson2012Gaussian, Luger2016EVEREST}, improving the characterization of signals \citep{Covino2020Looking, Kim2013STANDARDIZING, Barros2020Improvinga}, combining and improving RV datasets \citep{Farr2018Aldebaran, Aigrain2012Simple}, and more (see, for example, \citealt{Aigrain2023Gaussian}). Although GPs have seen some use in modeling stellar oscillations \citep{Barclay2015RADIAL, Grunblatt2017Seeing, Grunblatt2016K297b}, such efforts typically seek only to eliminate the stellar variability as a noise source.

An early application of GPs to model stellar oscillations was performed by \citet{Brewer2009Gaussian}, wherein individual modes of oscillation were modeled as independent stochastic processes with no direct mapping to the asteroseismic quantities. \citet{Foreman-Mackey2017Fast} demonstrated that with a suitable choice and parameterization of the kernels, solar-like oscillations can be modeled in the time domain and mapped directly to the asteroseismic quantities -- \rev{\numax\ (which measure the dominant envelope frequency of the excited oscillations) and \dnu\ (which measures the separation between modes of equal angular degree)}. Further work by \citet{Pereira2019Gaussian} used this approach to perform a systematic ensemble test of GP modeling applied to \tess-like artificial photometry, yielding results on par with or better than traditional frequency domain techniques. A significant barrier to the widespread adoption of this technique thus far has been the computational complexity of the models. However, recent advances in probabilistic programming models (e.g., \citealp{Foreman-Mackey2021Exoplaneta}) have significantly improved upon this, permitting high-dimensional models to be sampled efficiently.

In this paper, we present a time-domain model of solar-like pulsations, by modeling the light curve as a sum of stochastically excited, damped harmonic oscillators of varying quality factors. We show that our method presents a two-fold increase in accuracy compared to typical frequency domain methods for low SNR signals and for light curves with significant observational gaps. 
Our method is relatively fast for the \TESS\ data, provides robust uncertainties, and can easily be combined with other light curve models. To facilitate the interpretation of our GP analysis, we also provide a revised asteroseismic target list with computed probabilities of seismic detection for all stars brighter than 12th \tess\ magnitude, and demonstrate how GPs can accurately capture the variability in low signal-to-noise cases.

\section{The asteroseismic model}

\begin{figure}[t]
    \includegraphics[width=\linewidth]{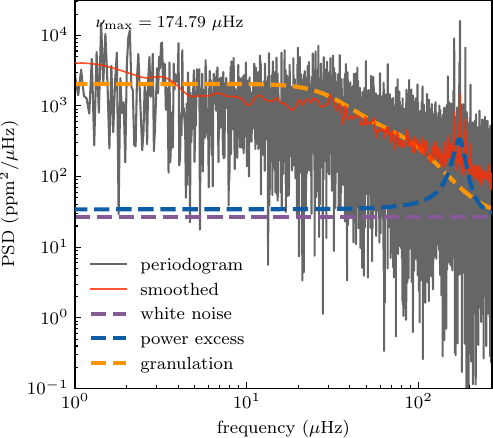}
    \includegraphics[width=\linewidth]{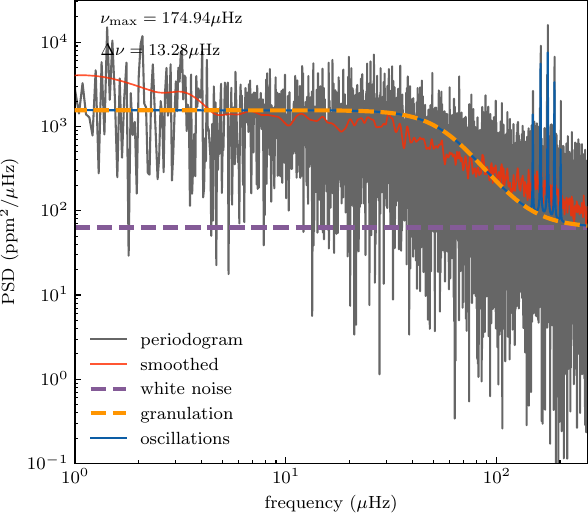}
    \caption{Top panel: the MAP solution for the power excess model. Bottom panel: the MAP solution for the oscillator comb model.}
    \label{fig:kic_example}
\end{figure}

\subsection{Theoretical framework}

We present here a slightly alternative formulation to describing these oscillations in the time-domain from \citet{Foreman-Mackey2018Scalable}. Oscillations in solar-like stars are stochastically excited by convection near the stellar surface, and naturally damped.  This process is well approximated by a damped simple harmonic oscillator, whose equation of motion is
\begin{equation}
    \Big[\frac{d^2}{dt^2} + \frac{\omega_0}{Q}\frac{d}{dt} + \omega_0^2\Big]y(t) = x(t).
    \label{eq:damped_sho}
\end{equation}
Here $\omega_0$ is the frequency of the undamped oscillator ($\omega_0 = 2 \pi f_0$), $Q$ is the quality factor (the damping time to period ratio), and $x(t)$ is a stochastic driving force. \cref{eq:damped_sho} describes a linear excitation forced by a stochastic function. In this case, stochastic motion is the combined effect of convective elements on the stellar surface. At late times, general solutions to \cref{eq:damped_sho} are described by convolution against an impulse-response function $h(t, t')$:
\begin{equation}
    \begin{aligned}
        y(t) & = \int_{-\infty}^{\infty} h(t,t')x(t') dt' \\&= \int_{-\infty}^t x(t') \cdot {1 \over \omega} \exp[- \eta (t - t')] \sin \left(\omega (t - t')\right).
    \end{aligned}
\end{equation}
By the convolution theorem, the power spectral density (PSD) of this process is given by the product of $S_0 = |\mathcal{F}[x]|^2$ with a Lorentzian transfer function (the Fourier transform of the impulse response):
\begin{equation}
    S(\omega) = \sqrt{\frac{2}{\pi}} \frac{S_0 \omega_0^4}{(\omega^2 - \omega_0^2)^2 + \omega_0^2 \omega^2 / Q^2}.
    \label{eq:psd}
\end{equation}
\citet{Foreman-Mackey2017Fast} remarks upon the behavior of \cref{eq:psd} in several interesting limiting regimes of $Q$.
For example, to describe the granulation components, one takes the overdamped limit of \cref{eq:damped_sho} to yield the Langevin equation
\begin{equation}
    \left[{\mathrm d^2 \over \mathrm d t^2} + \eta {\mathrm d \over \mathrm d t}\right] y = x(t),
\end{equation}
so that at late times
\begin{equation}
    \dot{y} = \int_{-\infty}^t x(t') e^{-\eta (t - t')} \mathrm d t'.
\end{equation}
When $x(t)$ is a white-noise process, $\left<x(t) x(s)\right> = S_0 \delta(t - s)$, the two-point correlation function of $\dot{y}$ is then
\begin{equation}
    \begin{aligned}
        \left<\dot{y}(s) \dot{y}(t)\right> & = S_0\int_{-\infty}^t \int_{-\infty}^s\ \mathrm d t_1 \mathrm d t_2  \delta(t_1 - t_2) e^{-\eta \left(t + s - t_1 - t_2\right)} \\
                                           & = {S_0 \over 2}\int_{-\infty}^{2\mathrm{min}(s, t)} e^{-\eta(s + t - u)} \mathrm{d} u = {S_0 \over 2 \eta} e^{- \eta |t - s|}.
    \end{aligned}
    \label{eq:overdamped}
\end{equation}
Correspondingly, the PSD of a granulation component is a semi-Lorentzian, or \citet{Harvey1985HighResolution}, profile:
\begin{equation}
    S(\omega) \propto {1 \over 1 + \left(\omega \over \eta\right)^2}.\label{eq:harvey}
\end{equation}
While the original construction of \cite{Harvey1985HighResolution} was intended for describing granulation in velocity (whence the examination of $\dot{y}$ rather than $y$), similar parameterizations are also used, or adapted, in photometric analysis. In such adaptations, the exponent of $\omega$ in the denominator of \autoref{eq:harvey} is often modified, or permitted to vary as a free parameter. For example, an index of 4 is used in modeling activity for transit photometry \citep{Dawson2014Large}. 

In modeling the power spectra of solar-like stars, a linear combination of such Harvey profiles is often used to describe the nonoscillatory stochastic variability, with each component corresponding to convective processes occurring at different timescales and amplitudes, such as mesogranulation and faculae. The question of how many such granulation components are appropriate to sufficiently fit observations remains open. \cite{Kallinger2014Connection} found that two granulation terms are usually sufficient to reproduce the granulation signal for \kepler\ data, whereas one term is sufficient for \tess\ data \citep{Barros2020Improvinga}. In this work, we run trials of both one and two granulation kernels separately.

Although the granulation timescale and power are well known to be proportional to \numax\ \citep[e.g.][]{Kallinger2014Connection}, we decouple these parameters from our model to ensure that there is no accidental sampling of an oscillation signal from unaccounted-for covariance with the granulation kernels. Additionally, in \citet{Pereira2019Gaussian}, the granulation terms are normalized such that they match the granulation term provided in \citet{Kallinger2014Connection}. We make no such normalization here, instead opting to use the natural parametrization of the kernel to improve sampling efficiency. After sampling, the parameters can be re-cast following eq.~12 of \citet{Pereira2019Gaussian}.

\subsection{The oscillator comb model}
\label{sec:combmodel}

Several studies have employed GPs to model the Gaussian envelope of oscillations. This power excess term is parameterized by \numax, the central frequency of the envelope. In practice, however, solar-like oscillators excite a range of modes within this envelope, which carry important information about fundamental stellar parameters. 
There are two approaches to model individual frequencies with GPs. One can either model each excited mode as an independent term (see, for example, \citealt{Brewer2009Gaussian}), or a parameterized model of the entire comb of frequencies \citep{Foreman-Mackey2017Fast}. In this paper, we follow the second option.

Mode frequencies of low-degree p-modes satisfy an asymptotic eigenvalue equation (e.g. \citealt{Bedding2006Solarlike}),
\begin{equation}
    \nu_{n, l} \sim \dnu\left(n + \frac{\ell}{2} + \epsilon_p\right) - \ell(\ell + 1)D_0,\label{eq:pmode}
\end{equation}
where $n$ and $\ell$ are integers describing the radial order and angular degree, respectively; \dnu\ is the large frequency separation between modes of equal angular degree; $D_0$ is a parameter sensitive to the sound-speed gradient near the core; and $\epsilon_p$, the p-mode phase offset, is sensitive to the structure of the star near the center and the surface \citep{Christensen-Dalsgaard2004Physics}. Mode frequencies of a given $\ell$ in the power spectrum can thus be approximated as a series of peaks, more or less equally spaced by $\dnu$.

Following \cite{Foreman-Mackey2017Fast}, we extend the GP model to accommodate multiple modes being excited simultaneously within an envelope of power excess. For each mode, indexed by an integer $j$, we parametrize the mode frequency as
\begin{equation}
    \omega_{0, j} = 2 \pi (\numax + j \dnu + \epsilon),\label{eq:wj}
\end{equation}
with amplitude
\begin{equation}
    S_{0, j} = \frac{A}{Q^2} \exp\left[- \frac{[j\dnu + \epsilon]^2}{2 W^2}\right],\label{eq:Sj}
\end{equation}
where $A$ and $W$ are nuisance parameters, and $\epsilon$ is a parameter that shifts the power excess envelope. In this paper, we generate a comb of frequencies for only a single angular degree (and assume that it is $\ell=0$). However, to accommodate additional angular degrees, one could extend the parametrization by introducing an additional term for the small separation $\delta\nu_{02}$, which is the approximate difference between the $\ell=0$ and $\ell=2$ modes:
\begin{equation}
    \omega_{2, j} = 2 \pi (\numax + j \dnu + \epsilon) - \delta \nu_{02}.
\end{equation}

We construct three models for the power spectrum: the power excess model, which only models the oscillation power excess envelope, the oscillator comb model, which models the entire comb of frequencies, and the `white noise' model, which fits only the granulation background and white noise level. The power excess model is parametrized by \numax\ and the amplitude of the power excess, $a_\mathrm{peak}$. The oscillator comb model is parametrized by \numax, \dnu, and the amplitude of the power excess, $a_\mathrm{peak}$. The oscillator comb model is more flexible than the power excess model, as it can model the entire comb of frequencies, and thus can be used to obtain \dnu\ directly. However, it is also more computationally expensive, as it requires the generation of many modes, each of which are represented by a simple harmonic oscillator kernel in the GP. The power excess model is more computationally efficient, but cannot be used to obtain \dnu. We further note the possibility of the oscillator comb model honing in on the incorrect sequence of degrees, thus reporting a measured value of half the actual \dnu. We avoid this in our models by pre-supplying an expected \dnu\ from the preliminary \numax\ following scaling relations of \citet{Stello2009relation}.  
\subsection{Regression}

\input{tables/priors.tex}

\begin{figure*}[t]
    \centering
    \includegraphics{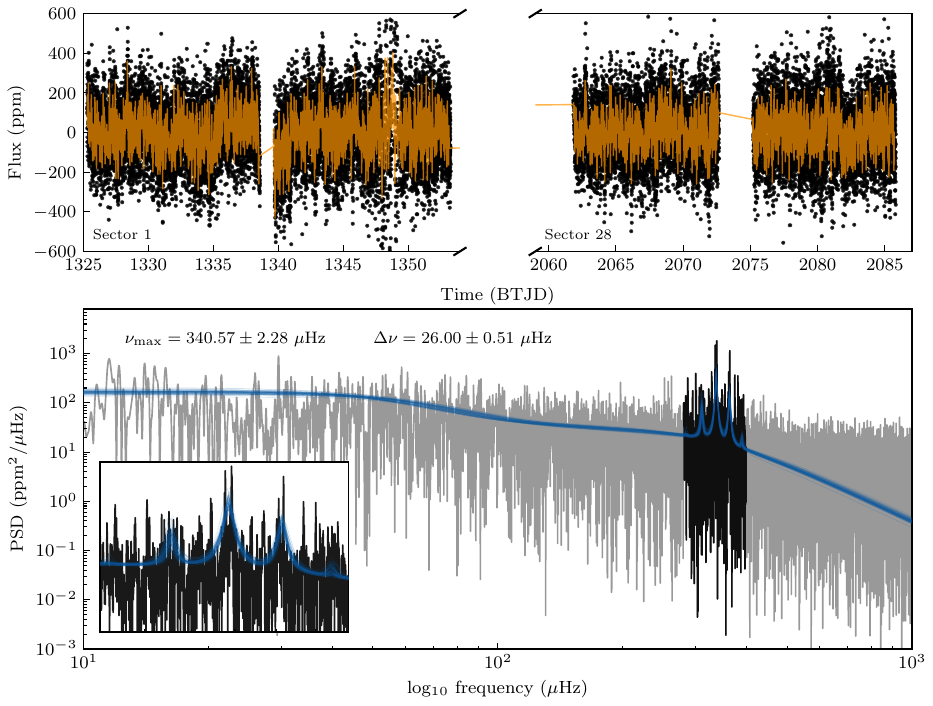}
    \caption{Model fit for \nuind. Top: The light curve of \nuind\ observed in Sector 1 and Sector 28, with the median posterior model light curve overplotted in orange. Bottom: The PSD of the combined light curve, with the median posterior model PSD overplotted in orange. The inset shows the power excess region.}
    \label{fig:nu_ind}
\end{figure*}

\begin{figure}[t]
    \includegraphics[width=\linewidth]{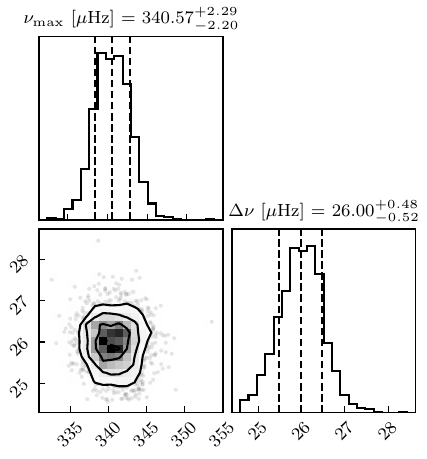}
    \caption{Posterior probabilities of \numax\ and \dnu\ for $\nu$~Ind.}
    \label{fig:nu_ind_corner}
\end{figure}

GP regression largely follows the same procedure as any other probabilistic model. The GP itself is described by a suitable two-point kernel function $k$ and mean function $\mu$, parameterized by $X$, chosen to accurately describe the data. The log-likelihood of the model is then
\begin{equation}
    \ln L(X) = -\frac{1}{2}\mathbf{r}^T \mathbf{K}^{-1}\mathbf{r} - \frac{1}{2} \ln|\det \mathbf{K}| - \frac{n}{2}\ln{2 \pi},
\end{equation}
where $\mathbf{r}$ is the vector of residuals after subtracting the mean function, $n$ is the number of data points, and the covariance matrix, $\mathbf{K}$, has matrix elements
\begin{equation}
    K_{ij}(X) = \sigma_i^2 \delta_{i,j} + k(t_i, t_j ;X).
\end{equation}
For each pair of points indexed by $i$ and $j$, $\sigma_i$ is the uncertainty in the observation of the $i$\textsuperscript{th} point (assumed to be uncorrelated between points, whence the appearance of the Kronecker delta ($\delta_{i,j}$), and $k(t_i, t_j; X)$ is the kernel, parameterized by $X$). Because the stellar variability is primarily stochastic, we may treat the GP as being stationary, so that $k(t_i, t_j ; X) = k(|t_i - t_j|; X)$ depends only on the absolute distance in time between points $i$ and $j$, and the mean function takes constant, or zero, value. Given this likelihood function and a set of suitable priors, posterior distributions on the model parameters $X$ may then be inferred in the usual Bayesian fashion.


We construct our models using \textsc{celerite2}, an updated version of the original \textsc{celerite} \citep{Foreman-Mackey2018Scalable} code designed to work with a range of modeling libraries in Python. We use the simple harmonic oscillator (SHO) kernel provided in \textsc{celerite2} within the Python interface, which maps directly to Eq.~\ref{eq:psd}. For the oscillator comb model (which incorporates \dnu), modes of equal angular degree are generated with a series of SHO kernels following Eqs.~\ref{eq:wj} and \ref{eq:Sj}. The \numax-only model uses a singular SHO kernel with a low quality factor to capture the power excess in the time domain. The granulation terms are overdamped SHO kernels, following Eq.~\ref{eq:overdamped}.

To determine the posterior distribution of the parameters we sample model parameters using the No-U-Turn Sampler (NUTS) implemented in the \textsc{PyMC3} package, used in several utilities and distributions from the \textsc{exoplanet} package \citep{Foreman-Mackey2021Exoplaneta}. NUTS provides some clear advantages over more widely used MCMC samplers in astronomical literature (e.g., \textsc{emcee}; \cite{Foreman-Mackey2013emcee}). One of these is the support for the differentiation of all components in the model, allowing for gradient-informed sampling. \rev{This process is more efficient than gradient-free inference, resulting in a greater number of effective samples. The initial position of the sampler is based on the Maximum A Posteriori (MAP) solution, which is obtained by optimizing the model using standard gradient descent routines. The priors of individual parameters in the model are outlined in \autoref{tab:priors}, with \dnu\ being scaled to \numax\ following \citep{Stello2009relation}. For the granulation parameters, the amplitudes are scaled to the standard deviation of the light curve, which is sufficiently close to the true value to properly capture the variability. The frequencies of the granulation parameters are wide normal distributions, centered over some typical values for red giants. A stricter prior could be chosen based on the work of \citet{Kallinger2014Connection} if desired. For the \numax\ model, the quality factor prior is low, describing a wide Gaussian envelope surrounding the power excess. The quality factor prior in the \dnu\ model is much higher because each individual mode is fit within some range. Unless otherwise stated, we sample all presented models for 2000 tuning and draw steps across two chains simultaneously, which allows comparisons of convergence rates. We primarily inspect the R-hat convergence diagnostic, which is a good indicator of convergence when $\hat{R} \sim 1$.}

As a test of the method, we apply our models to a simple example star KIC 11615890. The amplitudes of dipole oscillation modes in this star are heavily suppressed as a consequence of strong magnetic fields \citep{Stello2016Suppression}, greatly simplifying the application of our method here, as their absence simplifies the GP model required to describe it: the $\ell=0$ degree fit is sufficient to provide a good fit for both the \numax\ and \dnu\ models. We show the Maximum A Posteriori (MAP) optimized solutions for both the power excess envelope and oscillator comb model in Fig.~\ref{fig:kic_example}. The recovered \numax\ for the envelope and comb model is 174.79 and 174.94 $\mu$Hz respectively, with a $\Delta\nu$ of 13.28 $\mu$Hz.

\section{Time and frequency domain comparisons}

\subsection{The Subgiant $\nu$ Indi}

To demonstrate the GP model's resilience to gaps in data, which are typical of \tess\ observations, we apply our model to the well-known sub-giant oscillator, \nuind. \nuind\ was targeted for 120-s cadence observations by \tess, and shows a rich spectrum of solar-like oscillations at a \numax\ of around 350~$\mu$Hz \citep{Chaplin2020Age}. 
The power spectrum shows a series of radial ($\ell=0$), dipole ($\ell=1$), and quadrupole ($\ell=2$) modes in the \tess\ data. 
Due to its evolved status, its non-radial modes are not purely acoustic, instead showing ``mixed'' character from coupling to the internal g-mode cavity of the star \citep{Mosser2011Mixed}. Mixed-mode frequencies are more complicated to model, both evolutionarily --- as they change on much more rapid timescales than pure p-modes --- as well as in the power spectrum, as they do not satisfy the p-mode eigenvalue equation, \autoref{eq:pmode}. While it is straightforward to extend the oscillator comb to an arbitrary number of angular degrees, our chosen parameterization cannot describe such mixed modes. As such, we model only the $\ell=0$ radial modes following the procedure outlined in \autoref{sec:combmodel}. However, we also show that only fitting the radial modes is sufficient for obtaining useful measurements.

To emphasise the gaps in \tess\ photometry, we only fit the first and last available sectors of 120~s cadence data for \nuind\ (sectors 1 and 28), resulting in a light curve with an approximately 750~d gap. We fit both the \numax\ and \dnu\ models separately, optimize the model, and then run MCMC. For both models, we use two granulation kernels. For the power excess only model, we obtain \numax\ = $340\pm3~\mu$Hz, and for the oscillator comb model we get \numax\ = $340.5\pm2.1~\mu$Hz and \dnu\ = $25.97\pm0.51\mu$Hz. We show the results of the latter model in \autoref{fig:nu_ind}, and the posterior distribution of \numax\ and \dnu\ in Fig.~\ref{fig:nu_ind_corner}. These results are in excellent agreement with each other, and also with previous seismic investigations of $\nu$ Ind  \citep{Bedding2006Solarlike, Carrier2007Solarlike, Chaplin2020Age}. The distribution of the residuals is Gaussian, indicating a good model fit to our data. As can be seen in Fig.~\ref{fig:nu_ind}, the model includes only one sequence of frequencies ($\ell$=0) which do not directly align with the actual modes.

There is a strong covariance between \numax\ and $\epsilon$ in the oscillator comb model, arising from our prescription for separating equal-degree modes. By construction, \numax\ specifies the peak of some Gaussian window surrounding the power excess. Individual modes are then created around this value. Because the peak of the power excess does not necessarily align with the peak of the strongest mode, the $\epsilon$ value is required to offset the envelope to ensure that the modes align with the data. In practice, this results in a strong covariance between the parameters and a slight reduction in sampling efficiency. Additionally, since we are only modeling $\ell=0$ modes, the sampler can sometimes fail to converge due to $\ell=1$ and $\ell=2$ modes. This is avoided in most situation by supplying a sufficiently constrained prior for either \dnu\ or $\epsilon$.

\subsection{The Kepler red giant sample}

\begin{figure*}[t]
    \centering
    \includegraphics[width=\textwidth]{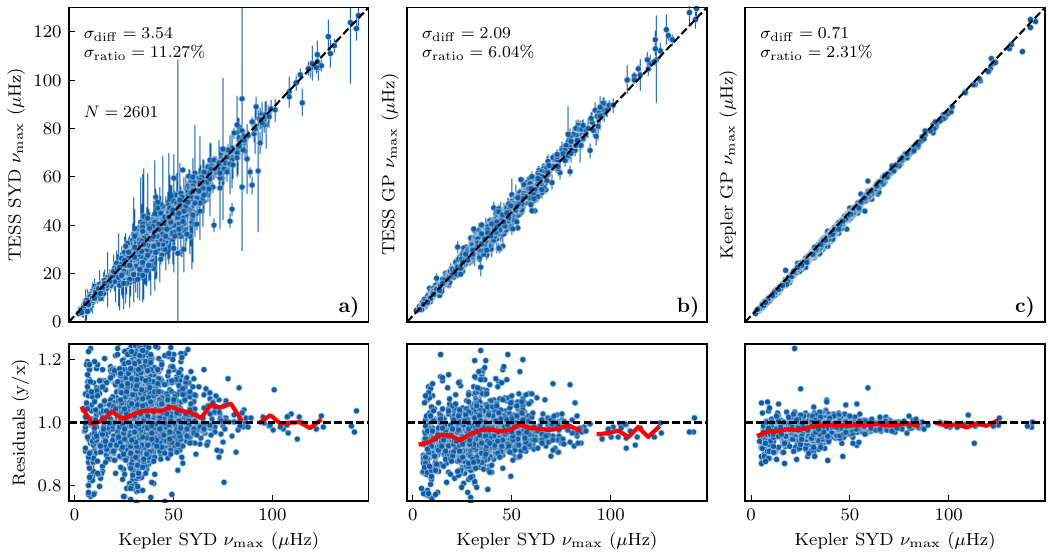}
    \caption{\rev{Comparison of the \kepler\ red giant sample observed by \tess. In each panel, the x-axis is the \kepler\ \numax\ measured by \citet{Yu2018Asteroseismology} in the frequency domain using the SYD pipeline. The axis labels follow the notation of <mission> <method>. \textbf{a)} Comparison of measured \numax\ from \kepler\ in the frequency domain (SYD) against the results of \cite{Stello2022TESS} \tess\ data measured in the frequency domain with the same method. \textbf{b)} The same compared against the GP model in the \tess\ data. \textbf{c)} The same compared against the GP model applied to the same \kepler\ data. \rev{Note that $\sigma_{\rm diff}$ and $\sigma_{\rm ratio}$ correspond to the standard deviation of the residual difference and ratio respectively.}}}
    \label{fig:numax_comparison}
\end{figure*}

We now seek to quantify the advantages of time domain asteroseismology for a larger sample. To do this, we measured \numax\ for the Kepler red giant sample as observed by \tess. This classification was originally performed by \cite{Stello2022TESS}, who manually searched for oscillations among the known \kepler\ red giant sample \citep{Yu2018Asteroseismology} observed by \tess in the original 30 minute cadence full-frame-images. They detect clear oscillations in around 2,655 stars using the SYD pipeline \citep{Huber2009Automated}. We use these results to validate our technique.

To ensure consistency across results, we use identical data and processing as described \citet{Stello2022TESS}. About half of the stars are observed for only 1 sector and the other half for two consecutive sectors, leading to many stars with a low oscillation SNR. For each star, we optimize the model on the light curve, and then sample for 1000 tuning and draw steps respectively initialized on the optimized solution. We record the complete MCMC chain and estimate convergence based on the r-hat statistic. For the final comparison sample, we keep only stars that pass the convergence criteria, resulting in a final sample of 2601 out of 2655 stars. The remaining stars were poorly initialized, such that the sampler did not have sufficient time to converge. Manual tweaking of the initial parameters for the failed stars leads to convergence.

To investigate any systematic offsets between the frequency and time-domain methods, we additionally run our sample on the original \kepler\ red giant light curves from \citet{Yu2018Asteroseismology}, which also used the SYD pipeline \citep{Huber2009Automated}. We fit only the \numax\ model in all cases.

We show the results in Fig.~\ref{fig:numax_comparison}. The spread in residuals for the GP model (Fig~4b) is significantly lower than that of the frequency domain method (Fig4a), \rev{indicating an 1.7 fold increase in accuracy}. We comment on three important features. First, the uncertainties on the GP results are significantly lower on average than that of the frequency domain method. Secondly, the residuals of the GP model are not centered around zero. This is to be expected; \numax\ summarizes the many individual mode amplitudes in solar-like oscillators, and various definitions for \numax\ exist. In the SYD pipeline used by \citet{Stello2022TESS}, \numax\ was defined as the location of the peak of the heavily smoothed background-corrected power spectrum. In the GP model, \numax\ is instead defined as the undamped frequency of a stochastically excited harmonic oscillator. This offset is especially apparent in \rev{Fig.~\ref{fig:numax_comparison},c where the frequency domain measurement of \numax\ in \kepler\ is compared against the GP measurement on the same light curve, indicating that the GP has a systematically larger estimated \numax\, on the order of 2 $\mu$Hz over the frequency domain methods. Finally, we note that the comparison of the \kepler\ data for the SYD and GP methods demonstrates functionally equivalent accuracy.}


\section{The revised asteroseismic target list}

\subsection{Background}

The \tess\ asteroseismic target list \citep[ATL,][]{Campante2016ASTEROSEISMIC,Schofield2019Asteroseismic} was constructed prior to the launch of \tess\ to optimize the selection of asteroseismic targets for 2-minute cadence observations. Given a set of fundamental stellar parameters, the ATL uses scaling relations to estimate oscillation amplitudes and combine these with estimates for the photometric noise to calculate a probability of detecting oscillations following \citet{Chaplin2011Ensemble}. The ATL has formed the basis for target selection for \tess\ guest investigator proposals by the \tess\ Asteroseismic Science Consortium for solar-like oscillators up until Cycle 6 (e.g., \tess\ GI G05155, G04104, G04220), and provides important information to evaluate the performance and occurrence of solar-like oscillations \citep{Hon2021Quick,Hatt2023Catalogue}.

Several new developments warrant a revision of the ATL. First, the original ATL used Tycho photometry \citep{Hog2000Tycho2} and Gaia DR2 parallaxes \citep{GaiaCollaboration2018Gaia} to calculate fundamental parameters. Tycho photometry becomes imprecise at $V\gtrsim 9$\,mag, thus leading to uncertain amplitude estimates that affect the detection probabilities. Furthermore, the ATL used only \teff\ and radius, while scaling relations for both amplitudes and \numax\ depend on surface gravity, which is now accessible through large-scale isochrone-fitting catalogs based on Gaia DR3 \citep{Collaboration2021Gaia}. Second, the newly available 20-second cadence photometry yields improved photometric precision \citep{Huber202220}, which is not captured in the original ATL. Finally, more sensitive detection techniques as described in Section 2 now allow for a more thorough evaluation of the \tess\ asteroseismic yield and the reliability of detection probabilities.

\subsection{Data}

\begin{figure}[t]
    \centering    
\includegraphics{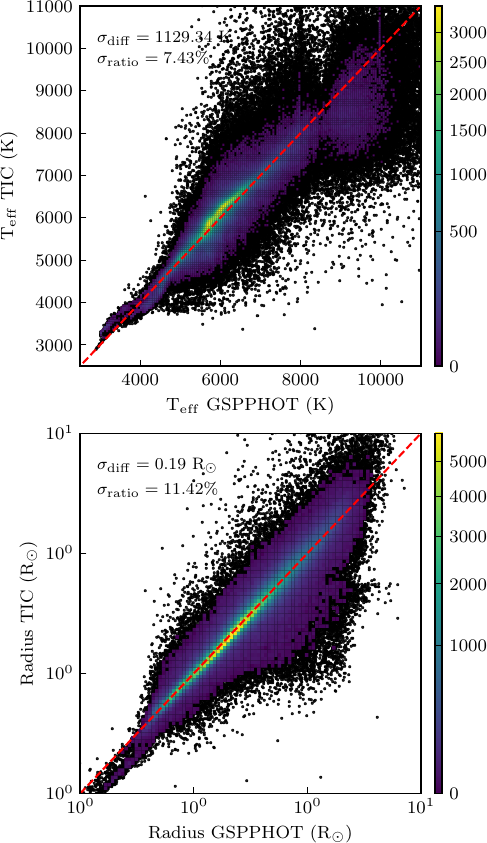}
    \caption{Comparison of Gaia \textsc{gspphot} values against the corresponding TIC value for a subset of 100,000 stars. The colorbar in both figures marks the count of stars in the bin. The red dashed line shows the 1:1 relation.}
\label{fig:tic_compare_gaia}
\end{figure}

The updated ATL uses surface gravity, effective temperature, stellar radius, \tess\ magnitude and coordinates as input parameters. Stellar parameters come from Gaia DR3 and the \tess\ Input Catalog (TIC) \citep{Creevey2023Gaia, Stassun2018TESS}. We start by selecting all stars from the \tess\ Input Catalog brighter than $T<12$\,mag that includes a pre-computed cross-match between TIC ID and Gaia DR2 ID, as well as surface gravities, temperatures, and radii. This sample consists of 5,138,580 stars. 

We then cross-match the Gaia DR2 to DR3 IDs using the best-neighbor conversion table provided by Gaia, from which we query the Gaia DR3 astrophysical parameters table. Gaia reports astrophysical parameters that are either derived from photometry and parallax (\textsc{gspphot}) or spectroscopy (\textsc{gspspec}). However, the coverage of astrophysical parameters from \textit{Gaia} is not complete; approximately 50\% of the sample have \textsc{gspphot} parameters, and 70\% have \textsc{gspspec} values. To supplement the sample we use of stellar parameters in the TIC, which are derived from photometry and Gaia DR2 parallaxes. Figure ~\ref{fig:tic_compare_gaia} shows a comparison \textsc{gspphot} and TIC parameters. We find good agreement for most stars, with a typical scatter of 10\% for temperature and 12\% for radius. A similar comparison with \textsc{gspspec} shows worse agreement.  This is consistent with the fact that the information from parallax and magnitude for \textsc{gspphot} parameters should provide a more accurate radius, while \logg\ is notoriously difficult to measure from spectral line broadening. As such, we prioritize values from DR3 \textsc{gspphot}, which provide a strong constraint on luminosity (or radius) through the parallax, followed by the TIC values, and use \textsc{gspspec} otherwise.


\subsection{Methodology}

\begin{figure}[t]
    \centering    
    \includegraphics{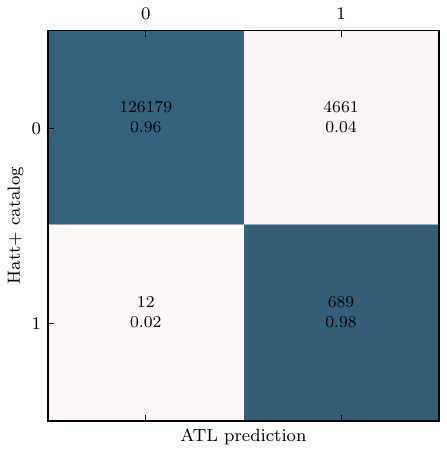}
    \caption{Comparison of detection probabilities of targeted stars against semi-automated asteroseismic detections provided in \citet{Hatt2023Catalogue} for stars whose predicted \numax\ are greater than 240 $\mu$Hz. Here, 0 and 1 denote non-detection and detection respectively, where a probability $>0.5$ implies detection in the ATL.}
\label{fig:confusion}
\end{figure}

We follow the basic methodology from the original ATL \citep{Chaplin2011Ensemble, Campante2016ASTEROSEISMIC, Schofield2019Asteroseismic}, with a few modifications. The original ATL used the following scaling relations for the frequency of maximum power (\numax) and oscillation amplitudes, which depend on only radius and temperature:

\begin{equation}
    \numax = \nu_{\rm max,\odot} \Big(\frac{R}{R_\odot}\Big)^{-1.85}\Big(\frac{T_{\rm eff}}{T_{\rm eff,\odot}}\Big)^{0.92},
\end{equation}

\begin{equation}
    A_{\rm oscillation} = \beta \Big(\frac{R}{R_\odot}\Big)^{1.85}\Big(\frac{T_{\rm eff}}{T_{\rm eff,\odot}}\Big)^{0.57},
\end{equation}
where $\beta$ is a scaling term to account for diminishing amplitudes approaching the red edge of the instability strip \citep{Schofield2019Asteroseismic};
\begin{equation}
    \beta = 1 - e^{\big[\dfrac{ (T_{\rm eff} - T_{\rm red \odot} L^{-0.093})}{1550 K}\big]}.
\end{equation}


Kepler observations have demonstrated that a more accurate scaling relation for oscillation amplitudes depends on mass, luminosity and temperature \citep{Huber2011Testing}:

\begin{equation}
    A_{\rm oscillation} = \frac{L^s}{M^t T_{\rm eff}^{r-1} c_K(T_{\rm eff})}
\end{equation}
where $s=0.838$, $t=1,32$, $r=2$, and 
\begin{equation}
    c_K = \Big(\frac{T_{\rm eff}}{5934 K} \Big)^{0.8}.
\end{equation}


We use the surface gravity and temperature to estimate \numax\ following the standard scaling relation \citep{Brown1991Detection}:
\begin{equation}
    \numax = \nu_{\rm max, \odot} \Big(\frac{10^{\log{g}}}{10^{4.44}}\Big) \Big(\frac{T_{\rm eff}}{T_{\rm eff \odot}}\Big)^{-0.5}
\end{equation}
where $g$ is surface gravity, and a standard temperature for the Sun of $T_{\rm eff \odot} = 5772$~K \citep{mamajekIAU2015Resolution2015}. We calculate luminosity using the standard Stefan-Boltzmann relation with radius and temperature, and obtain mass from the surface gravity and radius.


We further modify the predicted noise from \tess\ to account for the improved photometric precision of \tess\ 20\,second data. To do so, we take the calculated noise estimates from the original ATL function and multiply them by the scaling factors presented in table~1 of \citet{Huber202220a}. We use a simple linear interpolation scheme as implemented in \textsc{scipy} to fill in gaps \citep{Jones2001SciPy}. 

We generate four versions of the ATL: the 120s and 20s cadence probabilities for all stars in sectors up to the end of \tess\ Cycle 6, and the 120s and 20s cadence probabilities for all stars assuming that they have been observed in only 1 sector. These probabilities are calculated for stars that have been observed by the \tess\ cameras according to \textsc{tess-point}. The catalog of single sector probabilities serve as a resource for future target selection in upcoming \tess\ cycles. We also provide subsets of the complete 120s and 20s sample for stars that have been specifically targeted by \tess. The ATL sample is provided in Table~\ref{tab:ATL}.

\begin{table*}
\centering
\begin{tabular}{lrrrrrrrrrrrrr}
\toprule
 TIC & DR3 ID & T$_{\rm mag}$ & RA & Dec & Radius & T$_{\rm eff}$ & $\log{g}$ & $\nu_{\rm max}$ & P$_{\rm 120s}$ & $P_{\rm 20s}$  & P$_{\rm 120s, 1}$ & $P_{\rm 20s, 1}$\\
         &      &  [mag] &  &  & [$R_\odot$] & [K] & [dex] & [$\mu$Hz] &  &  \\
\midrule
 147993277 & 1062164554970946432 & 10.34 & 170.31 & 69.68 & 2.33 & 6211.01 & 3.91 & 874.24 & 0.01 & 0.01 & 0.01 & 0.01 \\
159504404 & 1071260535493824512 & 8.44 & 145.04 & 70.30 & 3.56 & 5915.60 & 3.50 & 348.51 & 0.35 & 0.48 & 0.05 & 0.07 \\
 280502586 & 5409350059063796992 & 10.55 & 144.32 & -48.62 & 1.74 & 8172.70 & 4.19 & 1450.53 & 0.01 & 0.01 & 0.01 & 0.01 \\
 124234835 & 6179156636869513472 & 7.77 & 197.25 & -34.35 & 3.55 & 7408.54 & 3.56 & 363.62 & 0.01 & 0.01 & 0.01 & 0.01 \\
 141306494 & 5179622467534734080 & 9.83 & 45.14 & -7.70 & 1.42 & 6118.76 & 4.13 & 1466.82 & 0.01 & 0.01 & 0.01 & 0.01 \\
  &&&&& \vdots\\
 379199238 & 1008455851496623872 & 10.66 & 93.67 & 64.66 & 1.33 & 5825.45 & 4.25 & 1997.31 & 0.01 & 0.01 & 0.01 & 0.01 \\
7982594 & 3782049416609809152 & 7.84 & 156.99 & -2.71 & 2.43 & 6414.70 & 3.85 & 753.05 & 0.03 & 0.03 & 0.02 & 0.02 \\
422590653 & 2647473405247080192 & 8.46 & 353.12 & 3.04 & 1.99 & 6557.56 & 3.95 & 935.70 & 0.01 & 0.02 & 0.01 & 0.01 \\
 70853984 & 283655037687700352 & 9.94 & 89.85 & 62.97 & 1.33 & 6225.55 & 4.15 & 1536.81 & 0.01 & 0.01 & 0.01 & 0.01 \\
274617294 & 1951734607913865856 & 7.90 & 321.75 & 36.75 & 1.90 & 6050.16 & 4.02 & 1152.73 & 0.04 & 0.06 & 0.02 & 0.02 \\
\bottomrule
\end{tabular}
\caption{A random subset of the revised ATL calculated in both 20s and 120s cadence. The columns P$_{\rm 120s, 1}$ and $P_{\rm 20s, 1}$ refer to the 120 and 20s probabilities calculated assuming the target was observed in only a single sector. The full version in electronic format is made available online.}
\label{tab:ATL}
\end{table*}

Following \citet{Schofield2019Asteroseismic}, we only report detection probabilities for stars with predicted $\numax > 240\mu$Hz (which means we exclude red giant stars). This is because the TIC does not supply \logg, and \textit{Gaia} surface gravities for giants were found to show significant scatter and systematic offsets compared to the asteroseismic sample from \citet{Huber2011Testing}. The lack of precise \logg\ values independent of asteroseismology is not surprising because stellar tracks converge and overlap around the giant branch, making it difficult to estimate mass (and thus \logg) from parallaxes, spectroscopy and photometry alone.

The calculated probabilities in the ATL are based on a fixed set of stellar parameters. For users wishing to input their own values (including for giant stars), we provide a Python package; \textsc{tess-atl}\footnote{\url{https://github.com/danhey/tess-atl}}, which exposes a command line interface to calculate detection probabilities with user-supplied stellar parameters. If no parameters are supplied, the target is automatically queried against the \tess\ input and Gaia DR3 catalogs, with the calculation being performed for an arbitrary input number of sectors. We supply this tool for target selection in future \tess\ cycles, as a way to easily measure the probability of variability detection given $N$ sectors in either 20 or 120-second cadence. We additionally provide the scripts used to generate the revised ATL, based on the functions available in \textsc{tess-atl}.

\subsection{Revised ATL Verification}

\begin{figure*}[t]
    \includegraphics[width=\linewidth]{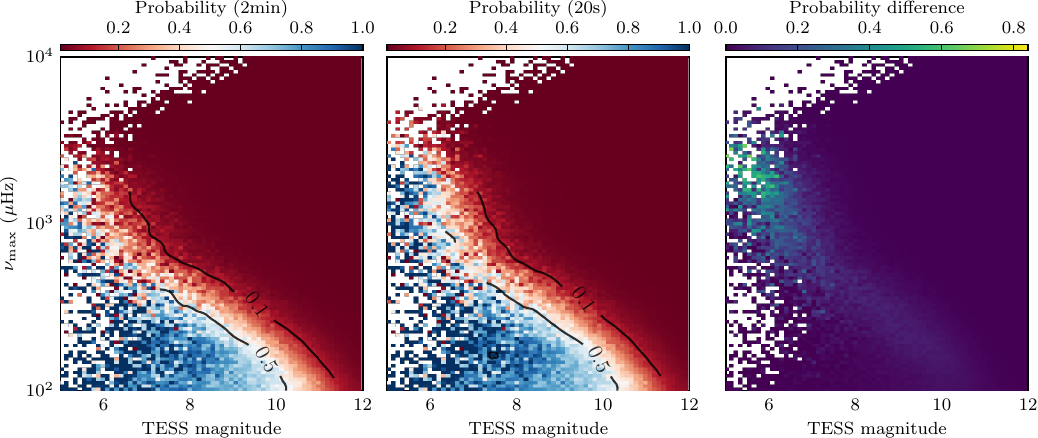}
    \caption{Detection probabilties from the revised ATL as a function of stellar magnitude (noise) and predicted \numax\ (signal), where the highest SNR is at the lower left corner and lowest SNR is at the upper right corner. In order from left to right are the 120 second cadence sample, the 20 second cadence sample, and their difference. The black lines mark smoothed contours of probability.}
    \label{fig:atl_numax}
\end{figure*}

\begin{figure}[t]
    \includegraphics[width=\linewidth]{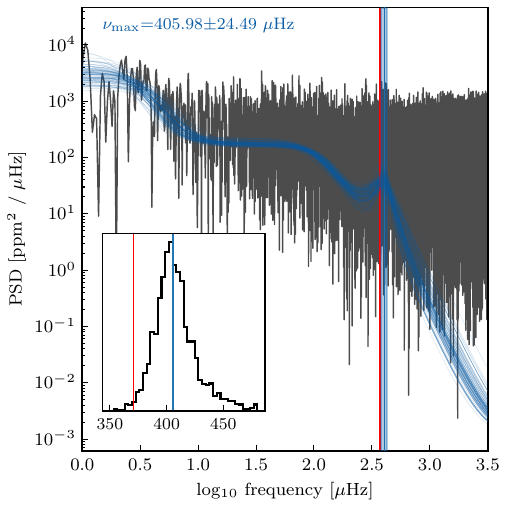}
    \caption{An example marginal detection in the \tess\ light curve of TIC 320382848. The blue lines show posterior samples from the trace, with the red vertical line indicating the ATL predicted \numax. The inset shows the posterior distribution of \numax\ samples.}
    \label{fig:marginal_detection}
\end{figure}

To test the revised ATL, we compare our predictions against the 120 second \tess\ detections from the \citet{Hatt2023Catalogue}, which involved a semi-automated classification using the official SPOC pipeline data. We limit this comparison to stars where the ATL predicts a \numax\ greater than 240 $\mu$Hz, and show our result in a confusion matrix in Fig.~\ref{fig:confusion}. Of their 701 detections, 689 (98\%) have a detection probability greater than 0.5 in our revised ATL. \citet{Hatt2023Catalogue} note that they expect a 7\% false positive rate, in good agreement with our predicted probabilities. However, the ATL tends to either over-estimate amplitudes or under-estimate noise levels based on the input stellar parameters, leading to an increase in false positives. We note that adjusting the probability cutoff can trade false positives for false negatives, and vice versa.

To investigate the accuracy, we inspected the power spectra of 100 randomly chosen light curves of stars with detection probabilities $>0.99$ that were not classified as detections by \citet{Hatt2023Catalogue}. Of these 100, 22 show clear by eye solar-like oscillations at the predicted \numax. We show an example of a low-amplitude detection in Fig.~\ref{fig:marginal_detection} using the GP method outlined in Section 2. The target, TIC 320382848, has a high predicted probability of detection but a low SNR. While the oscillations are not easily resolved by eye, we confidently obtain a \numax\ value close to what is predicted.

Other stars with high detection probabilities but no recorded detection either have unusual variations in their light curve (such as eclipses or rotation), or do not show any oscillation signal at all. We consider these missed detections to be due to the suppression of oscillation amplitudes with increased stellar activity \citep{Chaplin2011Predicting}, which is not taken into account in the ATL. 

Finally, we show the average detection probabilities of the 120 and 20-second cadence sample as a function of magnitude and \numax\ in Fig.~\ref{fig:atl_numax}. These panels show the expected trend of higher noise stars having a lower detection probability at fainter magnitudes. The panel of probability differences implies that 120-second cadence observations are sufficient for most of the bright, low \numax\ \tess\ targets. The 20-second cadence noise properties become increasingly favorable for the brightest, highest \numax\ oscillators -- especially for the small island of bright, high frequency stars observed by \tess\ which have high detection probabilities. 

\section{Conclusion}

In this paper, we have developed and expanded upon the work of \citet{Foreman-Mackey2017Fast} to fit time-domain asteroseismic models to \tess\ photometry. We have also provided a revised target list for solar-like oscillators. Our main conclusion are as follows:

\begin{itemize}
    \item We construct a Gaussian Process framework to fit time-domain asteroseismic signals. This includes a model to fit only the dominant envelope of power excess (\numax), and a model to fit individual modes of equal angular degree (\numax + \dnu). We show that these models are  robust against gaps in data and variations in cadence that is typical of \tess\ and ground-based data.
    \item We demonstrate that the \kepler\ \numax\ values obtained with the Gaussian Process model are nearly twice as precise as traditional frequency domain methods, using \kepler\ photometry as the ground truth. Because the speed of Gaussian Processes scales inversely with the length of the data-set, this method is almost uniquely suited to the short, relatively sparse \tess\ and ground-based data.
    \item We calculate updated asteroseismic detection probabilities for \tess\ based on stellar parameters from Gaia DR3 and the \tess\ Input Catalog. We calculate this new asteroseismic target list (ATL) for all main-sequence and subgiant stars with a \numax\ greater than 240 $\mu$Hz brighter than 12th magnitude, with separate probabilities for 20 and 120 second cadence. We also provide a software tool to calculate asteroseismic detection probabilities for any target of interest and any number of sectors.
    \item  Using an example of a star with significant detection probabilities that has not been reported in the literature, we demonstrate that our GP framework is capable of reliably extracting low signal-to-noise oscillations. The difficulty of detecting low SNR oscillations with traditional frequency-based methods may explain the current reduced yield of detections compared to pre-flight expectations. 

\end{itemize}

The results presented here demonstrate the potential of time-domain asteroseismology for the detection and characterization of solar-like oscillations in the \tess\ data. Combining the new ATL with our method to the archival \tess\ data and upcoming new observations has the potential to significantly increase the current asteroseismic yields from the \tess\ mission. The GP method also has strong potential for or ground-based photometric surveys, which typically obtain sparse observations, such as ATLAS \citep{Tonry2018ATLAS}, ASAS-SN \citep{Kochanek2017AllSky}, ZTF \citep{Bellm2014Zwicky}.

\section*{acknowledgements}

D.R.H.\ acknowledges the support of the National Science Foundation(AST-2009828). D.H. acknowledges support from the Alfred P. Sloan Foundation, the National Aeronautics and Space Administration (80NSSC21K0652, 80NSSC22K0303, 80NSSC23K0434, 80NSSC23K0435), and the Australian Research Council (FT200100871).

\vspace{5mm}



\begin{singlespace}
    \bibliographystyle{aasjournal}
    \bibliography{library}
\end{singlespace}

\end{document}

%% file: commands.tex
\usepackage{xspace,amsmath,amssymb}
\usepackage{xcolor, fontawesome}
\usepackage{tabularx}
\usepackage[LGR,T1]{fontenc}
\newcommand{\textgreek}[1]{\begingroup\fontencoding{LGR}\selectfont#1\endgroup}
\usepackage[utf8]{inputenc}
\usepackage{textcomp} 
\usepackage[capitalize]{cleveref}
\definecolor{linkcolor}{rgb}{0.1216,0.4667,0.7059}
\definecolor{xlinkcolor}{cmyk}{1,1,0,0}

\newcommand{\numax}{\mbox{$\nu_{\rm max}$}}

\newcommand{\kepler}{{\em Kepler}}

\newcommand{\dnu}{\mbox{$\Delta{\nu}$}}

\newcommand{\teff}{\ensuremath{T_\mathrm{eff}}}

\newcommand{\logg}{\ensuremath{\mathrm{log}\ g}}

\newcommand{\TESS}{\mbox{\textit{TESS}}\xspace}
\newcommand{\tess}{\mbox{\textit{TESS}}\xspace}
\newcommand{\nuind}{\textgreek{n}~Indi\xspace}

%% file: authors.tex
\correspondingauthor{Daniel Hey}
\email{dhey@hawaii.edu}
\author[0000-0002-0786-7307]{Daniel Hey}
\affiliation{Institute for Astronomy, University of Hawai`i, 2680 Woodlawn Drive, Honolulu, HI 96822, USA}

\author[0000-0001-8832-4488]{Daniel Huber}
\affiliation{Institute for Astronomy, University of Hawai`i, 2680 Woodlawn Drive, Honolulu, HI 96822, USA}
\affiliation{Sydney Institute for Astronomy (SIfA), School of Physics, University of Sydney, NSW 2006, Australia}

\author[0000-0001-7664-648X]{Joel Ong}
\altaffiliation{Hubble Fellow}
\affiliation{Institute for Astronomy, University of Hawai`i, 2680 Woodlawn Drive, Honolulu, HI 96822, USA}

\author[0000-0002-4879-3519]{Dennis Stello}
\affiliation{Sydney Institute for Astronomy, School of Physics, University of Sydney, NSW 2006, Australia}
\affiliation{School of Physics, UNSW Sydney, NSW 2052, Australia }
    
\author[0000-0002-9328-5652]{Daniel Foreman-Mackey}
\affiliation{Center for Computational Astrophysics, Flatiron Institute, New York, NY 10010, USA}

%% file: tables/priors.tex

\begin{table}
    \caption{Priors used in the models.}

    \begin{equation*}
        \begin{aligned}
            \text{Noise and Offsets: } & \left\{
            \begin{array}{rl}
                \sigma/\text{ppm} & \sim \mathcal{N}(0, 10) \\
                \mu/\text{ppm}    & \sim \mathcal{N}(\bar{y}, 1) \\
            \end{array}\right.                                                                 \\
            \text{Granulation: }       & \left\{
            \begin{array}{rl}
                a_{\rm gran, 1}/\text{ppm}      & \sim \ln(\mathcal{N}(\sigma^2_\mathrm{y}, 10)) \\
                a_{\rm gran, 2}/\text{ppm}      & \sim \ln(\mathcal{N}(\sigma^2_\mathrm{y}, 10)) \\
                \omega_{0, 1} / \mathrm{d^{-1}} & \sim \ln(\mathcal{N}(1, 10)) \\
                \omega_{0, 2} / \mathrm{d^{-1}} & \sim \ln(\mathcal{N}(5, 10)) \\
            \end{array}\right. \\
            \text{Power Excess: }      & \left\{
            \begin{array}{rl}
                \numax / \mathrm{\mu Hz}   & \sim \ln(\mathcal{N}(\nu_{\rm max, init}, 100))                          \\
                a_\mathrm{peak}/\text{ppm} & \sim \ln(\mathcal{N}(\sigma^2_\mathrm{y}, 10)) \\
                Q                          & \sim \ln(\mathcal{N}(1, 5))
            \end{array}\right.                        \\    
            \text{Oscillator Comb: }   & \left\{
            \begin{array}{rl}
                \Delta\nu / \mathrm{\mu Hz} & \sim \ln(\mathcal{N}(\Delta\nu_{\rm init}, 5)) \\
                a_\mathrm{peak}/\text{ppm}  & \sim \ln(\mathcal{N}(1, 5)) \\
                Q                           & \sim \ln(\mathcal{N}(5, 2)) \\
                \epsilon / \mathrm{\mu Hz}  & \sim \ln(\mathcal{N}(0, 1)) 
            \end{array}\right.
        \end{aligned}
    \end{equation*}
    {\footnotesize
    \textsc{Notes} --- $\sigma$ and $\mu$ are the white noise and mean flux level of the light curve respectively. $\mathcal{N}(\mu, \sigma)$ is the normal distribution of mean $\mu$ and standard deviation $\sigma$. $y$ refers to the flux of the light curve.
    }
    \label{tab:priors}
\end{table}